\documentclass[letterpaper, 10 pt, conference]{ieeeconf} 

\IEEEoverridecommandlockouts                             
\usepackage{amsmath} 
\usepackage{amssymb}  
\usepackage{textcomp}

\newtheorem{myrem}{Remark}

\usepackage{graphicx}
\usepackage{textcomp}
\usepackage{float}
\usepackage{cite}

\usepackage{booktabs}
\usepackage{multirow}

\usepackage{algorithm}
\usepackage{algpseudocode}
\algrenewcommand\algorithmicrequire{\textbf{Input:}}
\algrenewcommand\algorithmicensure{\textbf{Output:}}

\usepackage[table]{xcolor}

\usepackage{subcaption}

\newcommand*{\QEDA}{\hfill\ensuremath{\blacksquare}}

\usepackage{color}

\title{\LARGE \bf
Rollout-Based Charging Scheduling for Electric Truck Fleets 

\vspace{1.5pt} 
in Large Transportation Networks
}

\author{Ting Bai, Xinfeng Ru, Shaoyuan Li, and Andreas A. Malikopoulos
\thanks{This research was supported in part by NSF under Grants CNS-2401007, CMMI-2348381, IIS-2415478, and in part by MathWorks.}\vspace{1.5pt}
\thanks{T. Bai and S. Li are with the School of Automation and Intelligent Sensing, Shanghai Jiao Tong University, Shanghai 200240, China. E-mails: \{{\tt\small tingbai, syli\}@sjtu.edu.cn}}
\thanks{X. Ru is with the Department of Electrical and Electronic Engineering, East China University of Science and Technology, Shanghai, 200240, China. E-mail: {\tt\small xfru@ecust.edu.cn}}
\thanks{A. A. Malikopoulos is with the Information and Decision Science Lab, School of Civil $\&$ Environmental Engineering, Cornell University, Ithaca, New York, U.S.A. E-mail: {\tt\small amaliko@cornell.edu}}
}
\begin{document}

\maketitle
\thispagestyle{empty}
\pagestyle{empty}

\begin{abstract}
In this paper, we investigate the charging scheduling optimization problem for large electric truck fleets operating with dedicated charging infrastructure. A central coordinator jointly determines the charging sequence and power allocation of each truck to minimize the total operational cost of the fleet. The problem is inherently combinatorial and nonlinear due to the coupling between discrete sequencing decisions and continuous charging control, rendering exact optimization intractable for real-time implementation. To address this challenge, we propose a rollout-based dynamic programming framework built upon an inner–outer two-layer structure, which decouples ordering decisions from the schedule optimization, thus enabling efficient policy evaluation and approximation. The proposed method achieves near-optimal solutions with polynomial-time complexity and adapts to dynamic arrivals and time-varying electricity prices. Simulation studies show that the rollout-based approach significantly outperforms conventional heuristics with high computational efficiency, demonstrating its effectiveness and practical applicability for real-time charging management in large-scale transportation networks.
\end{abstract}

\section{Introduction}\label{Section I}
Road freight transportation is a major contributor to carbon emissions, accounting for over $7\%$ of global CO$_2$ emissions and projected to rise to approximately $16\%$ by 2050~\cite{raoofi2024system}. Against this backdrop, reducing the environmental impact of freight transport has become a critical priority for achieving carbon neutrality targets. Among various mitigation strategies, such as truck platooning~\cite{10209062,mahbub2023_automatica,9993403}, alternative fuels (e.g., hydrogen and biofuels), and logistics optimization, the electrification of road freight has emerged as a more direct and promising solution. Compared with conventional diesel-powered vehicles, electric trucks (ETs) offer multiple advantages, including reduced greenhouse gas emissions, lower cost of ownership, and quieter operation.

As a clean mode of transportation, however, the large-scale deployment of ET fleets is hindered by several challenges~\cite{zhao2024challenges,morganti2018technical}, among which limited driving range is a primary concern. Despite ongoing advancements in battery technology, current ET battery capacities are typically insufficient to sustain long-haul delivery missions, forcing trucks to stop and recharge their batteries midway. In addition, charging infrastructure remains scarce~\cite{6919255,10590837} and the charging process is time-consuming~\cite{bakker2025strategic,zhang2018optimal}. These factors introduce additional operational uncertainties for fleet electrification, including the risk of congestion at charging stations and potential violations of delivery deadlines, thus undermining timely delivery and efficient fleet management.

Charging at dedicated stations provides a viable pathway to alleviate drivers’ anxiety in locating suitable charging facilities and to reduce queuing risks arising from uncoordinated charging behaviors. However, when multiple trucks charge at the same station, they inevitably compete for limited charging resources~\cite{wu2023electric}. This competition necessitates the development of effective charging scheduling strategies that meet the charging demands of individual trucks while enabling reliable station operations~\cite{liu2020optimal}. To facilitate the wide adoption of ETs in road freight transport, this paper studies the charging scheduling optimization problem for ET fleets, where trucks share a dedicated charging station with limited charging ports and power capacity. The objective is to optimally schedule the charging process of each truck so as to minimize the total operational cost of the fleet. 

Over the past few years, charging scheduling for electric vehicles has been extensively studied from both station-centric and vehicle-centric perspectives. Station-based approaches coordinate charging activities to optimize system-level objectives, such as infrastructure utilization, power balancing, and operational cost~\cite{elghitani2020efficient,zhang2021distributed,gupta2020collaborative}. However, these works often overlook the heterogeneous costs and operational requirements of individual vehicles. In contrast, vehicle-based methods enable decentralized decision-making, where vehicles determine their own charging plans to minimize travel or waiting time~\cite{qin2011charging,zhang2013charging}, while respecting regulation constraints~\cite{10147895}. Nevertheless, the lack of coordination with stations may result in inefficient resource utilization and increased operational uncertainty. More recently, learning-based approaches have been proposed to handle stochastic environments and complex system dynamics~\cite{wan2018model,li2019constrained,zhao2024reinforcement}. Yet, these approaches lack explicit modeling of resource competition or operational constraints, such as limited charging ports and power capacity. While our recent work in~\cite{10932686} considers realistic operational constraints, it relies on a simplified first-come-first-served policy to capture interactions between trucks and charging stations, without incorporating charging sequence optimization within the scheduling framework.

In this paper, we study the charging scheduling optimization problem for large-scale ET fleets with a dedicated charging station subject to operational constraints, including limited charging ports, station-level power capacity, and time-varying electricity prices. A central coordinator jointly determines the charging schedules and service sequence of all trucks within the fleet with the objective of minimizing the total operational cost. The problem is inherently combinatorial, as the number of feasible service orders grows factorially with the fleet size, rendering exact solutions computationally intractable. To overcome this challenge, we present an inner-outer structure that separates sequencing decisions from schedule optimization. This two-layer modeling framework enables a rollout-based approach to efficiently approximate the optimal policy while balancing solution quality and computational tractability. The main contributions of this paper are summarized as follows:
\vspace{1pt}

\begin{itemize}
\item [\small{$\bullet$}] We formulate the charging scheduling problem as a mixed-integer nonlinear program, which captures both the discrete ordering decisions and continuous charging optimization under practical operational constraints. 
\vspace{3pt}

\item [\small{$\bullet$}] We present an inner-outer two-layer optimization structure that decomposes the problem into an outer-layer ordering problem and an inner-layer scheduling problem, enabling efficient evaluation of candidate solutions.
\vspace{3pt}

\item [\small{$\bullet$}] We develop a rollout-based dynamic programming (DP) approach that achieves a trade-off between solution quality and computational efficiency, with provable policy improvement and polynomial-time complexity.
\end{itemize}

Simulation studies based on realistic data demonstrate the effectiveness and superior performance of our approach.  

The rest of the paper is organized as follows. Section~\ref{Section II} presents the problem formulation. Section~\ref{Section III} introduces the two-layer optimization structure and the rollout-based DP approach. Section~\ref{Section IV} analyzes the optimality and computational complexity of the proposed approach. Section~\ref{Section V} provides simulation results, followed by concluding remarks and future research directions in Section~\ref{Section VI}.

\section{Problem Formulation}\label{Section II}
This section formulates the charging scheduling optimization problem. We start by modeling the charging process under limited charging facilities. Next, we present the constraints associated with the charging assignment and charging order. Finally, we define the control objective and formulate the charging scheduling problem. 

\subsection{Modeling of the Charging Process}
We consider a large-scale ET fleet served by a dedicated charging station, where the number of available charging ports is denoted as $C\!\in\!\mathbb{Z}_{+}$, with $\mathbb{Z}_+$ representing the set of positive integers. The set of trucks requiring charging at the station is denoted by $\mathcal{N}\!:=\!\{1,\dots,N\}$. For each truck $i\!\in\!\mathcal{N}$, the information provided to the coordinator (i.e., fleet manager) includes its arrival time $t_i^a$, remaining battery upon arrival $E_i$, charging demand $\Delta{E}_i$, and the latest departure time $d_i$, which is determined by the delivery deadline. The battery capacity of the truck is denoted as $E_i^{\max}$. 

We first introduce the model of the charging process. For each truck $i\!\in\!\mathcal{N}$, let $P_i^{\max}$ be its maximum admissible charging power. The charging power available at each port $c\!\in\!\{1,\dots,C\}$ is denoted as $P_c$. To describe the assignment between trucks and charging ports, we define a binary variable $a_{i,c}(t)\!\in\!\{0,1\}$, which equals $1$ if truck $i$ is connected to port $c$ at time $t$, and $a_{i,c}(t)\!=\!0$ otherwise. The effective charging power of truck $i$ at time~$t$ is then denoted as
\begin{align}
P_i(t)=\sum_{c=1}^{C} a_{i,c}(t)\cdot\min\!\big(P_i^{\max},P_c\big).\label{Equ.1}
\end{align}

To meet the charging demand of truck $i$, we have
\begin{align}
\Delta E_i=\int_{t_i^{s}}^{t_i^{d}}\!P_i(t)dt,\label{Equ.2}
\end{align}
where $t_i^{s}$ and $t_i^{d}$ denote the start and end charging times of truck $i$, respectively. Accordingly, the charging duration of truck $i$ is represented as
\begin{align}
T_i=t_i^{d}-t_i^{s}.\label{Equ.3}
\end{align}

\subsection{Ordering Constraints}
In this subsection, we introduce the constraints induced by the charging assignment and ordering decisions. Let $\Omega$ denote the set of all feasible \emph{per-port orderings}, where each element consists of $C$ ordered sequences that partition the set $\mathcal{N}$. For each charging port $c\!\in\!\{1,2,\dots,C\}$, let
\begin{align}
\omega_c=\{i_1^c, i_2^c, \dots, i_{|\omega_c|}^c\} \label{Equ.4}
\end{align}
be the ordered sequence of trucks assigned to port $c$, where $|\omega_c|$ is the number of trucks in $\omega_c$. 

The collection of ordering sequences across all charging ports is then given by
\begin{align}
\omega = \{\omega_1, \omega_2, \dots, \omega_C\}\!\in\!\Omega. \label{Equ.5}
\end{align}
Here, we note that, given a charging order $\omega$, the assignment of trucks to charging ports is implicitly determined. That is, the assignment variable $a_{i,c}(t)$ can be represented as
\begin{align}
a_{i,c}(t)=
\begin{cases}
1, & \text{if } i\!\in\!\omega_c \text{ and } t\!\in\![t_i^{s}, t_i^{d}],\\
0, & \text{otherwise.}\nonumber
\end{cases}
\end{align}

For each $\omega$, the following holds:
\begin{subequations}\label{Equ.6}
\begin{align}
&\bigcup_{c=1}^C \omega_c = \mathcal{N}, \label{Equ.6a}\\
&\omega_{c} \cap \omega_m = \emptyset, \quad \forall c\neq{m}. \label{Equ.6b}
\end{align}
\end{subequations}
The constraints in \eqref{Equ.6} ensure that each truck $i\!\in\!\mathcal{N}$ is assigned to exactly one charging port, and that all trucks are scheduled. To prevent overlapping charging operations on the same port, we define the \emph{precedence arc set} induced by $\omega\!\in\!\Omega$ as
\begin{align}
\mathcal{A}(\omega) := \bigcup_{c=1}^C \Big\{ (i_{\ell}^{c}, i_{\ell+1}^{c}) : \ell = 1,2,\dots,|\omega_{c}| - 1 \Big\}, \label{Equ.7}
\end{align}
where each arc $(i,j)\!\in\!\mathcal{A}(\omega)$ indicates that truck $i$ is served before truck $j$ on the same charging port. Thus, these precedence relations enforce a sequential charging order on each port, ensuring that no two trucks are charged simultaneously on the same port.

Furthermore, to enforce non-overlapping charging operations along each port, we impose the following precedence constraints:
\begin{align}
t_j^{s}\geq t_i^{d}, \quad \forall (i,j)\!\in\!\mathcal{A}(\omega),
\end{align}
which guarantee that, for each precedence arc $(i,j)\!\in\!\mathcal{A}(\omega)$, the charging of truck $j$ can only start after the charging of truck~$i$ has been completed.

\subsection{Charging Scheduling Optimization Problem}
To achieve efficient and cost-effective operation of the ET fleet, the central coordinator optimizes the charging schedule of all trucks. This involves determining the optimal charging order, i.e., the sequence in which trucks are assigned to charging ports, the corresponding start charging times and charging powers. The objective is to minimize the system-wide operational cost, including the total waiting loss, energy costs, and penalties incurred from delivery delays.

Next, we formulate the charging scheduling optimization problem. For each truck $i \!\in\!\mathcal{N}$, let $\epsilon_i$ and $\gamma_i$ denote the per-unit-time waiting cost and the penalty for delivery deadline violations, respectively. The electricity price is time-varying and denoted as $\lambda(t)$. In addition, the station-level power capacity is denoted by $P^{\max}$, which represents the maximum aggregate charging power provided by the station.

Given the set of trucks $\mathcal{N}$, with arrival times $\{t_i^a\}_{i\in\mathcal{N}}$, remaining batteries $\{E_i\}_{i\in\mathcal{N}}$, charging demands $\{\Delta{E}_i\}_{i\in\mathcal{N}}$, and the latest departure times $\{d_i\}_{i\in\mathcal{N}}$, the central coordinator determines the charging schedule by addressing the following optimization problem: 
\begin{subequations}\label{Problem}
\begin{align}
\!\!\!\min_{\substack{\omega\in\Omega,\\ \{t_i^s,P_i(t)\}_{i=1}^N}}
&  \sum_{i=1}^N\!\Bigg(\!
\int_{t_i^{s}}^{t_i^{d}}\!\lambda(t)P_i(t)dt+\epsilon_i(t_i^{s}\!-\!t_i^a)\!\nonumber\\
&\qquad\qquad\qquad\quad +\gamma_i\max\big\{t_i^{d}\!-\!d_i,0\big\}\!\Bigg),\label{Problem.a}\\
\!\!\!\!\mathrm{s.\ t.} \ \ 
&\int_{t_i^{s}}^{t_i^{d}}\!\!P_i(t)dt=\Delta E_i, \quad \forall i\in\mathcal{N}, \label{Problem.b}\\
&E_i+\Delta E_i\leq E_i^{\max}, \quad \forall i\!\in\!\mathcal{N}, \label{Problem.c}\\
&0\leq P_i(t)\leq P_i^{\max}, \quad \forall i\!\in\!\mathcal{N},\ \forall t\!\in\![t_i^{s},t_i^{d}],\label{Problem.d}\\
&\sum_{i=1}^{N}P_i(t)\le P^{\max}, \quad \forall t,\label{Problem.e}\\
&t_i^{s}\ge t_i^a,\ \ t_i^{d}\ge t_i^{s}, \quad \forall i\!\in\!\mathcal{N}, \label{Problem.f}\\
&t_j^{s}\ge t_i^{d}, \quad \forall (i,j)\!\in\!\mathcal{A}(\omega),\label{Problem.g}
\end{align}
\end{subequations}
where, as defined earlier, $\mathcal{A}(\omega)$ is the precedence arc set and $\omega$ denotes the charging ordering.

The charging scheduling problem formulated in \eqref{Problem} is a \emph{mixed-integer nonlinear program} (MINLP) involving discrete and continuous decision variables. The discrete component arises from the ordering variable $\omega\!\in\!\Omega$, which encodes the assignment and sequencing of trucks to the limited number of charging ports. The continuous component consists of the charging powers $P_i(t)$ and the start charging time $t_i^{s}$, which are coupled through integral constraints and time-dependent electricity costs. The joint optimization over $\omega$ and the continuous variables results in a nonconvex and combinatorial problem. As the fleet size $N$ grows, the number of feasible orderings increases factorially, making exact optimization intractable for large-scale systems.

\section{Rollout-Based Charging Scheduling Approach}\label{Section III}
In this section, we develop an efficient rollout-based DP approach for solving the charging scheduling problem. We first reformulate the problem into a two-layer optimization framework. Based on this reformulation, we present the rollout-based DP method and the implementation algorithm.

\subsection{Two-Layer Optimization Structure}
The problem formulated in \eqref{Problem} involves both discrete and continuous decision variables. Specifically, the charging order $\omega\!\in\!\Omega$ is combinatorial, whereas the charging schedule $\{P_i(t), t_i^{\rm s}\}_{i \in \mathcal{N}}$ is continuous. To address the original problem efficiently, we decompose it into a two-layer optimization framework consisting of an outer layer and an inner layer:
\begin{itemize}
    \item \textbf{Outer layer (ordering problem):} This layer determines the charging order $\omega\!\in\!\Omega$, which specifies the assignment of trucks to charging ports and their service sequence.
    
    \item \textbf{Inner layer (scheduling problem):} For a given ordering $\omega$, this layer optimizes the charging schedule $u\!:=\!\{P_i(t), t_i^{\rm s}\}_{i \in \mathcal{N}}$ to minimize the objective function subject to the operational constraints.
\end{itemize}

Mathematically, the outer layer solves the problem
\begin{align}
\min_{\omega\in\Omega}\ J(\omega),\label{Equ.10}
\end{align}
where $J(\omega)$ is the optimal value of the inner-layer problem defined as
\begin{align}
\!\!J(\omega) :=\!\! \min_{u \in \mathcal{U}(\omega)}  & \sum_{i=1}^N \Bigg(\!
\int_{t_i^{s}}^{t_i^{d}}\!\lambda(t) P_i(t)dt
+ \epsilon_i (t_i^{s} - t_i^a) \nonumber\\
& \qquad\qquad\qquad
+ \gamma_i \max\{t_i^{d} - d_i, 0\}\!\Bigg), \label{Equ.11}
\\
\mathrm{s.\ t.} &\ \ \ \ \ \mathrm{\eqref{Problem.b}}-\mathrm{\eqref{Problem.g}},\nonumber
\end{align}
where $\mathcal{U}(\omega)$ denotes the feasible set of continuous decision variables for a given ordering $\omega$.
\vspace{2pt}

\begin{myrem}\label{Remark1}
Under the two-layer decomposition, the inner layer solves a nonlinear continuous optimization problem, while the outer layer addresses a combinatorial optimization over all feasible orderings. As the size of $\Omega$ grows factorially with $N$, enumerating all possible orderings $\omega$ becomes computationally intractable, especially for large-scale ET fleets.
\end{myrem}

\subsection{Rollout-Based DP Approach}
Building upon the two-layer optimization framework, we now introduce the rollout-based DP approach. The outer-layer ordering problem is interpreted as a finite-horizon sequential decision process. Specifically, let $k\!\in\!\{0,\dots,N\!-\!1\}$ denote the decision stage, where at each stage one truck is assigned to a charging position, i.e., appended to one of the port sequences. The system state at stage $k$ is denoted by $x_k$ and defined as
\begin{align}
x_k := \tilde{\omega}_k := \{\tilde{\omega}_k^1,\dots,\tilde{\omega}_k^C\},\label{Equ.12}
\end{align}
where $\tilde{\omega}_k$ denotes a partially constructed charging order across all ports, satisfying $\bigcup_{c=1}^C \tilde{\omega}_k^{c}\!\subseteq\!\mathcal{N}$. Accordingly, the set of unassigned trucks at stage $k$ is expressed as
\begin{align}
\mathcal{N}_k^{\mathrm{U}} = \mathcal{N} \setminus \bigcup_{c=1}^C \tilde{\omega}_k^c. \label{Equ.13}
\end{align}

To proceed, we denote by $U_k(x_k)$ the action set at state $x_k$, where each action $u_k\!\in\!U_k(x_k)$ represents assigning one unassigned truck to a specific charging port $c$, namely,
\begin{align}
u_k := (i,c), \quad i\!\in\!\mathcal{N}_k^{\mathrm{U}},\ c\!\in\!\{1,\dots,C\}.\nonumber
\end{align}
After applying $u_k$, the system state evolves according to 
\begin{align}
x_{k+1} = f_k(x_k,u_k) = \tilde{\omega}_{k+1} = \{\tilde{\omega}_{k+1}^1,\dots,\tilde{\omega}_{k+1}^C\},\label{Equ.14}
\end{align}
where
\begin{align}
\tilde{\omega}_{k+1}^m =
\begin{cases}
\tilde{\omega}_k^m, & \text{if } m \neq c,\\
\tilde{\omega}_k^c \circ i, & \text{if } m = c,
\end{cases}
\label{Equ.15}
\end{align}
with $\tilde{\omega}_k^c\circ{i}$ denoting appending truck $i$ to the end of the sequence $\tilde{\omega}_k^c$.

Based on the sequential decision formulation below, we define the performance of a policy and characterize the optimal solution. For any complete ordering $\omega\!\in\!\Omega$, the total cost is denoted as $J(\omega)$, as given in \eqref{Equ.11}. Under a policy $\pi\!=\!\{\mu_0,\mu_1,\dots,\mu_{N-1}\}$, a complete ordering $\omega_\pi\!\in\!\Omega$ is generated sequentially starting from the initial state $x_0$, where at each stage the action $\mu_k(x_k)$ assigns one unassigned truck to a charging port. The cost-to-go associated with policy $\pi$ is then defined as
\begin{align}
J_\pi(x_0) := J(\omega_\pi),\nonumber
\end{align}
where $\omega_\pi$ denotes the complete charging order induced by~$\pi$. By construction, $\omega_\pi\!\in\!\Omega$ satisfies the ordering constraints, and the cost $J(\omega_\pi)$ is obtained by solving the inner-layer optimization problem, thus ensuring feasibility with respect to all scheduling constraints. Accordingly, the optimal policy is defined as
\begin{align}
\pi^* \in \arg\min_{\pi \in \Pi} J_\pi(x_0), \label{Equ.16}
\end{align}
where $\Pi$ denotes the set of admissible policies that generate feasible orderings in $\Omega$.

\subsection{Rollout Policy and Algorithm Design}
As discussed in Remark~\ref{Remark1}, the state space grows factorially with $N$, making the exact solution of problem~\eqref{Equ.16} computationally prohibitive. To address this challenge, we next develop a rollout-based approximation approach for the charging scheduling problem.

Let $\bar{\pi}$ denote a given \emph{base policy} that generates a feasible ordering (e.g., earliest-deadline-first). The rollout policy $\tilde{\pi}$ improves upon $\bar{\pi}$ through a one-step lookahead optimization. At each state $x_k$, the rollout decision is given by
\begin{align}
\tilde{\mu}_k(x_k) \in \arg\min_{u_k \in U_k(x_k)}
\Big\{ \hat{J}\big(f_k(x_k,u_k); \bar{\pi}\big) \Big\}, \label{Equ.17}
\end{align}
where $\hat{J}(f_k(x_k,u_k);\bar{\pi})$ denotes the approximate cost-to-go defined as
\begin{align}
\hat{J}\big(f_k(x_k,u_k);\bar{\pi}\big):= J\!\left(\tilde{\omega}(x_k,u_k;\bar{\pi})\right), \label{Equ.18}
\end{align}
where $\tilde{\omega}(x_k,u_k;\bar{\pi})\!\in\!\Omega$ denotes the complete charging order obtained by applying action $u_k$ at state $x_k$, while assigning all remaining trucks according to the base policy $\bar{\pi}$. That is, each candidate action is evaluated by constructing a complete ordering via one-step lookahead and computing its associated cost through the inner-layer optimization problem~\eqref{Equ.11}.

Given the above formulation, the complete rollout-based procedure for solving the charging scheduling problem is summarized in Algorithm~\ref{Algorithm 1}. 

It is worth noting that the developed rollout-based approach decouples the combinatorial ordering decisions from the continuous charging optimization, thereby yielding a tractable approximation to the original MINLP. The resulting per-stage computational complexity is polynomial (as we will discuss below), making our method suitable for large-scale fleet operations where exact DP or global MINLP approaches are computationally prohibitive.

\begin{algorithm}[t]
\caption{Rollout-Based Approach for the Two-Layer Charging Scheduling Problem}
\label{Algorithm 1}
\begin{algorithmic}[1]
\Require Parameters $\{t_i^a,E_i,d_i,\Delta E_i,E_i^{\max},P_i^{\max},\epsilon_i,\gamma_i\}_{i\in\mathcal{N}}$, station parameters $\lambda(t)$, $C$, $P^{\max}$, base policy $\bar{\pi}$.
\Ensure Rollout-based charging order $\omega^{\mathrm{RO}}$ and charging schedules $\{P_i(t),t_i^{s}\}_{i\in\mathcal{N}}$.

\State \textbf{Outer-layer initialization:} 
\State Set the stage index $k \gets 0$
\State Initialize the state
$x_k \gets x_0\!:=\!\{\emptyset,\dots,\emptyset\}$

\While{$\bigcup_{c=1}^C x_k^c \neq \mathcal{N}$}
    \State Compute the set of unassigned trucks
    $\mathcal{N}_k^{\mathrm{U}}$ by \eqref{Equ.13} 
    \State Construct the action set
    \[
    U_k(x_k) \gets \{(i,c): i\!\in\!\mathcal{N}_k^{\mathrm{U}},\ c\!\in\!\{1,\dots,C\}\}
    \]
    \State Initialize $\hat{J}^* \gets +\infty$

    \For{each $u_k\!=\!(i,c)\!\in\! U_k(x_k)$}
        \State Compute the next state $x_{k+1}$ by~\eqref{Equ.14} and \eqref{Equ.15}
        \State Obtain the complete ordering $\tilde{\omega}(x_k,u_k;\bar{\pi})$
        \State \textbf{Inner-layer evaluation:} solve problem \eqref{Equ.11} under $\tilde{\omega}(x_k,u_k;\bar{\pi})$, subject to constraints \eqref{Problem.b}--\eqref{Problem.g}
        \State Set $
        \hat{J}\big(f_k(x_k,u_k);\bar{\pi}\big)
        \gets
        J\big(\tilde{\omega}(x_k,u_k;\bar{\pi})\big)
        $
        \If{$\hat{J}\big(f_k(x_k,u_k);\bar{\pi}\big) < \hat{J}^*$}
            \State $\hat{J}^* \gets \hat{J}\big(f_k(x_k,u_k);\bar{\pi}\big)$
            \State $u_k^* \gets u_k$
        \EndIf
    \EndFor

    \State Apply $u_k^*$ and update the state $
    x_{k+1} \gets f_k(x_k,u_k^*)$
    \State Set $k \gets k\!+\!1$ and $x_k \gets x_{k+1}$
\EndWhile

\State Set the rollout-based charging order $\omega^{\mathrm{RO}} \gets x_k$
\State Obtain the charging schedules $\{P_i(t),t_i^s\}_{i\in\mathcal{N}}$ by solving problem~\eqref{Equ.11} for the ordering $\omega^{\mathrm{RO}}$ 
\end{algorithmic}
\end{algorithm}

\section{Optimality and Computational Complexity}\label{Section IV}
This section presents an analysis of the proposed rollout-based charging scheduling approach. We first investigate its optimality properties, followed by a computational complexity analysis to demonstrate its efficiency.

\subsection{Discussion on Optimality}
The rollout policy $\tilde{\pi}$ improves upon the base policy $\bar{\pi}$ by incorporating a one-step lookahead mechanism that accounts for both the immediate cost and its impact on future decisions. Under standard assumptions, the rollout policy satisfies the well-known improvement property~\cite{bertsekas1997rollout,bertsekas2021rollout}
\begin{align}
J_{\tilde{\pi}}(x_0) \le J_{\bar{\pi}}(x_0),\label{Equ.19}
\end{align}
indicating that it performs no worse than the base policy. This property follows from the fact that, at each decision stage, the rollout policy explicitly evaluates the cost associated with each candidate action using the base policy as a benchmark for future decisions. As a result, $\tilde{\pi}$ can be viewed as a policy improvement step over $\bar{\pi}$. 
\vspace{2pt}

\begin{myrem}\label{Remark 2}
The performance of the rollout-based solution critically depends on the choice of the base policy, as indicated by \eqref{Equ.19}. In particular, a stronger base policy provides a more accurate approximation of the cost-to-go, leading to improved decision quality at each stage and, consequently, better overall performance of the rollout policy. 
\end{myrem}

In practice, the rollout-based approach often achieves near-optimal performance while maintaining significantly lower computational complexity compared to exact DP or global optimization methods. Within the proposed two-layer optimization framework, each candidate action is evaluated through the inner-layer optimization problem, which inherently respects all problem constraints, ensuring that the resulting solution is both feasible and cost-efficient.

\subsection{Complexity Analysis}
To demonstrate the computational efficiency of the rollout-based solution approach, in this subsection, we analyze the complexity of Algorithm~\ref{Algorithm 1}.

At each decision stage $k\!\in\!\{0,\dots,N\!-\!1\}$, the algorithm constructs the action set
\begin{align}
U_k(x_k) = \{(i,c): i\!\in\!\mathcal{N}_k^{\mathrm{U}},\ c\!\in\!\{1,\dots,C\}\},\nonumber
\end{align}
whose cardinality is denoted as
\begin{align}
|U_k(x_k)| = C(N\!-\! k),\nonumber
\end{align}
where $|\mathcal{N}_k^{\mathrm{U}}|\!=\!N\!-\!k$ represents the number of unassigned trucks at stage $k$.

For each action $u_k\!\in\!U_k(x_k)$, the rollout procedure performs a one-step lookahead evaluation, which involves: (i) generating the next state, (ii) completing the partial ordering using the base policy, and (iii) solving the inner-layer optimization problem \eqref{Equ.11} to evaluate the total cost. Among these steps, the dominant computational cost arises from solving the inner problem. Let $T_{\mathrm{inner}}(N)$ denote the computational complexity of solving the inner-layer optimization problem for a complete ordering of size $N$. Then, the computational cost at stage $k$ is
\begin{align}
T_k\!=\!O\big(C(N\!-\!k)\!\cdot\!T_{\mathrm{inner}}(N)\big).\label{Equ.20}
\end{align}

Summing over all stages, we obtain the total computational complexity
\begin{align}
T_{\mathrm{rollout}} 
= \sum_{k=0}^{N-1} T_k
= O\!\left(C N^2\!\cdot\!T_{\mathrm{inner}}(N)\right).\label{Equ.21}
\end{align}

In general settings where the charging power is time-varying, the inner-layer problem corresponds to a continuous optimization problem with time-dependent variables and coupling constraints. If this problem admits a polynomial-time solution, its complexity can be expressed as
\begin{align}
T_{\mathrm{inner}}(N) = O(N^\alpha),\nonumber
\end{align}
for some $\alpha\!>\!0$. Here, the parameter $\alpha$ characterizes the computational complexity of the inner-layer problem as a function of the fleet size $N$. In this paper, the inner-layer problem is typically a convex optimization problem when the charging dynamics and cost functions are convex, in which case efficient polynomial-time algorithms (e.g., interior-point methods) can be applied, leading to a moderate value of $\alpha$ (e.g., $\alpha\!\approx\!2$--$3$). If nonlinear or nonconvex charging models are considered, the increased coupling among decision variables and constraints may lead to higher computational complexity, resulting in a larger effective value of $\alpha$.

Consequently, the overall complexity of the rollout-based approach in Algorithm~\ref{Algorithm 1} is denoted as
\begin{align}
T_{\mathrm{rollout}} = O\!\left(C N^{2+\alpha}\right).\label{Equ.22}
\end{align}

The above analysis shows that the overall computational complexity of the rollout-based approach scales polynomially with the fleet size $N$. 

\begin{myrem}\label{Remark 3}
Compared with exact DP, whose complexity grows exponentially with $N$ due to the enumeration of all possible orderings, the proposed rollout-based method significantly reduces the computational burden by limiting the search to one-step lookahead decisions. As a result, the developed approach achieves a favorable trade-off between computational efficiency and solution quality.
\end{myrem}
\section{Simulation Studies}\label{Section V}
This section evaluates the performance of the proposed rollout-based approach using realistic truck data. We first describe the simulation setup. Subsequently, we present the simulation results and comparisons with exact and heuristic methods. The code implementation is available online.\footnote{See code implementation at: {https://github.com/kth-tingbai/Rollout-based-Charging-Schedule-Optimization.git}}

\subsection{Setup}
\begin{table}[t]
\centering
\caption{Time-of-Use Electricity Prices.}
\label{Table I}
\fontsize{9}{11}\selectfont
\renewcommand{\arraystretch}{1}
\begin{tabular*}{\columnwidth}{@{\extracolsep{\fill}}lcc}
\hline
\hline
Label & Time Period & Price [€/kWh] \\
\hline
Off-peak & 00:00 -- 06:00 & 0.101 \\
Morning peak & 06:00 -- 09:00 & 0.174 \\
Mid-peak & 09:00 -- 12:00 & 0.128 \\
Daytime off-peak & 12:00 -- 17:00 & 0.110 \\
Evening peak & 17:00 -- 21:00 & 0.202 \\
Off-peak & 21:00 -- 24:00 & 0.101 \\
\hline
\hline
\end{tabular*}
\end{table}

The parameters of the ETs are set based on publicly available data for heavy-duty ETs developed by Scania~\cite{ElectricTruck}. Specifically, we consider a fleet of $N$ trucks, each with a battery capacity of $E_i^{\max}\!\!=\!468$ kWh and a maximum charging power of $P_i^{\max}\!=\!350$ kW. The initial battery energy $E_i$ of each truck is randomly generated within $[20\%, 80\%]$ of its capacity. We assume that each ET is charged to full capacity; thus, the charging demand is given by $\Delta E_i\!=\!E_i^{\max}\!-\!E_i$. The arrival times $t_i^a$ are randomly sampled within a predefined time window.

The charging deadlines $d_i$ are determined based on delivery schedules with allowable slack to capture operational flexibility, given by
\begin{align}
d_i = t_i^a + \frac{S \cdot \Delta E_i}{P_i^{\max}}, \label{Equ.23}
\end{align}
where $S\!\ge\!1$ is a slack parameter. The waiting cost and tardiness penalty coefficients are set to balance service efficiency and deadline adherence, with $\epsilon_i\!=\!2$ \texteuro/min and $\gamma_i\!=\!10$ \texteuro/min. The time-varying electricity price $\lambda(t)$ follows a time-of-use tariff, as specified in Table~\ref{Table I}. In addition, the charging power at each port is set to $P_c\!\in\!\{300,350\}$ kW.

To construct the rollout policy, we consider the following three heuristic methods as base policies:
\begin{itemize}
\item \textbf{FCFS} (First-Come, First-Served): trucks are prioritized according to their arrival times;
\item \textbf{EDF} (Earliest Deadline First): trucks are prioritized based on their deadlines;
\item \textbf{SCDF} (Smallest Charging Demand First): trucks are prioritized according to their charging demands.
\end{itemize}

In the following, we use RO (FCFS), RO (EDF), and RO (SCDF) to denote the rollout-based solutions constructed using FCFS, EDF, and SCDF as the base policies, respectively.

\subsection{Comparison Results}
\subsubsection{Comparison with the Exact Solution}

\begin{figure}[t]
    \centering
     \includegraphics[width=0.98\linewidth]{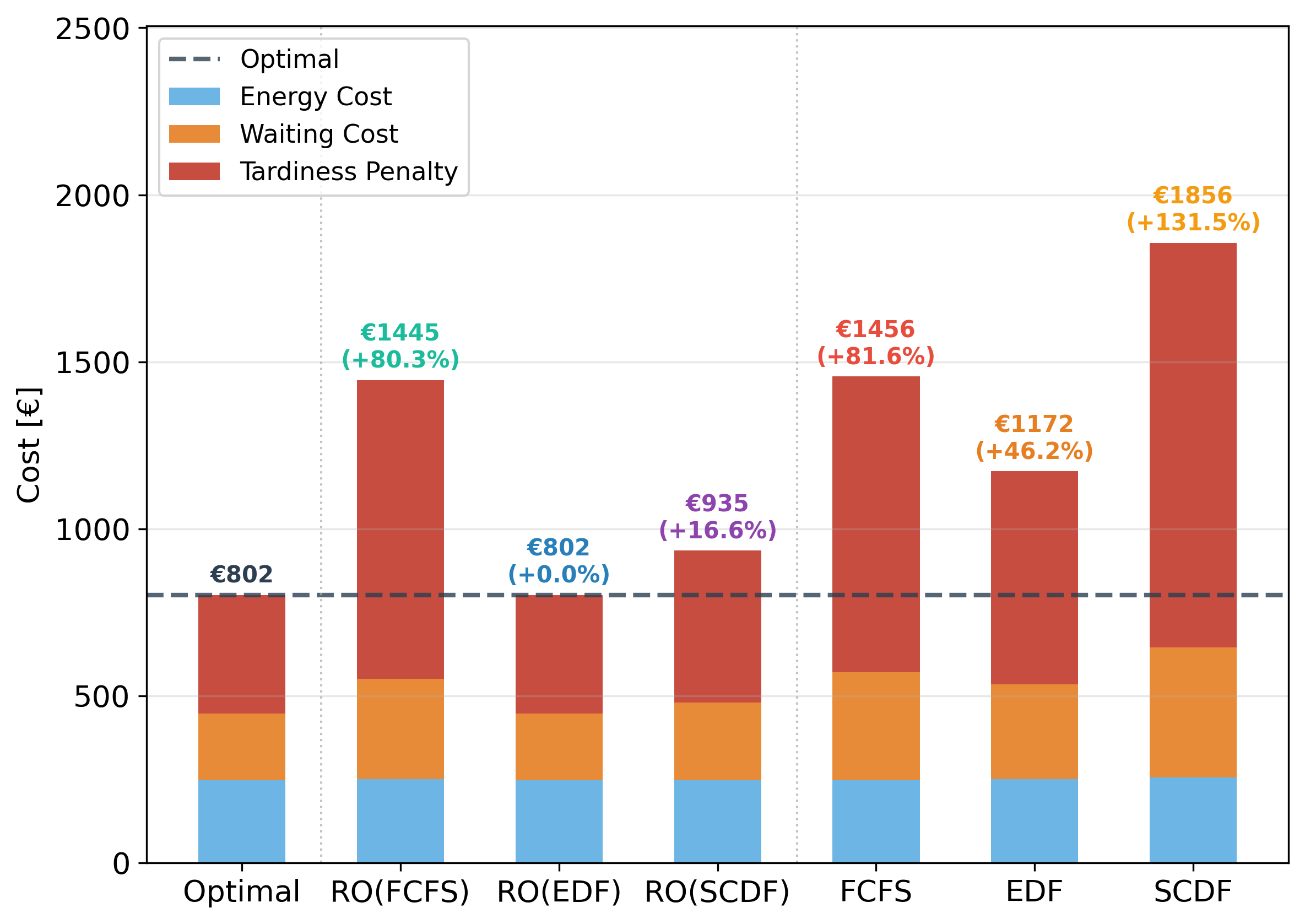}
      \caption{Comparison of the total cost ($N\!=\!8$).}
      \label{Fig.2}
\end{figure}

\begin{figure}[t]
\centering
\begin{subfigure}[t]{0.48\textwidth}
    \centering
    \includegraphics[width=\linewidth]{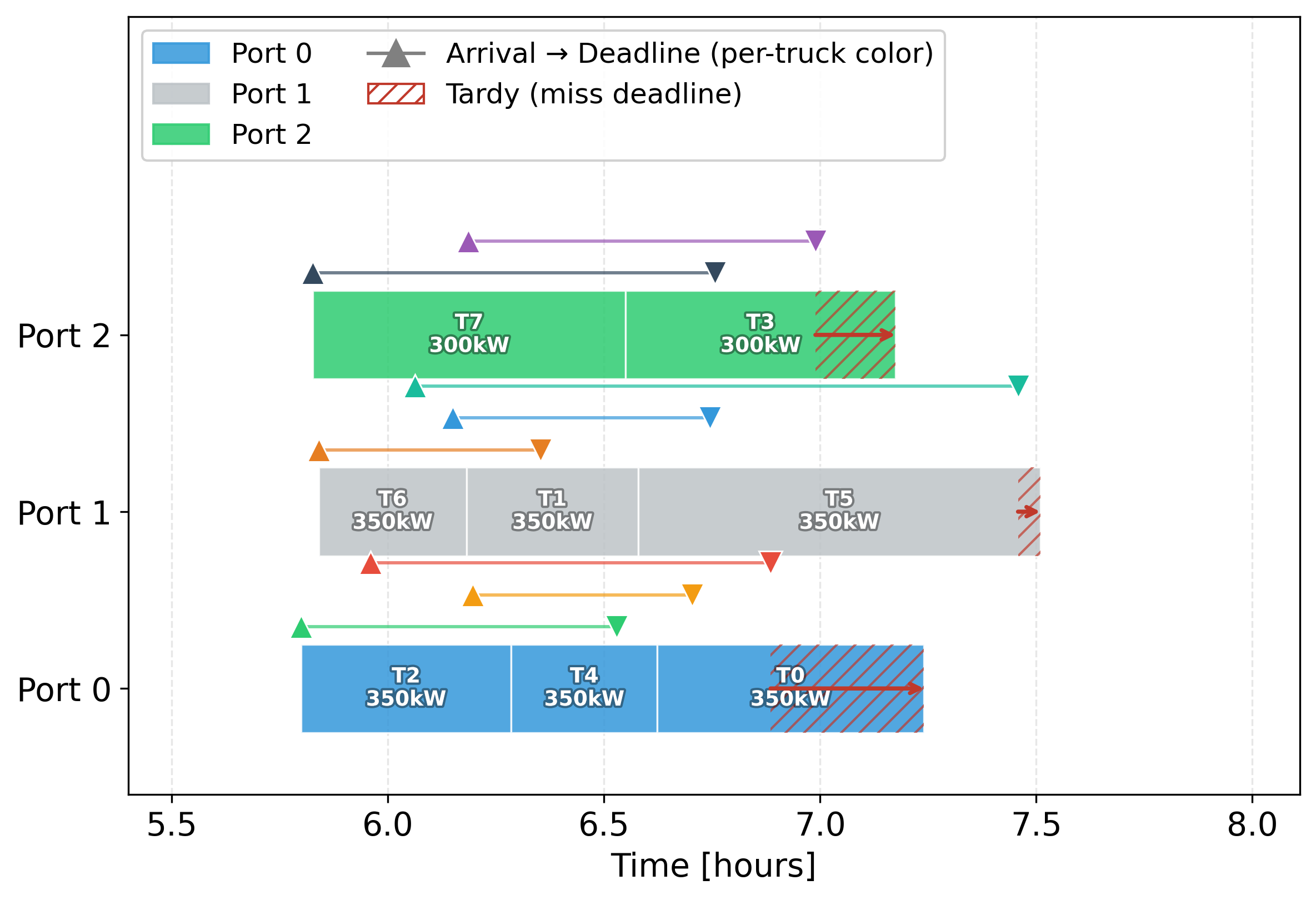}
    \caption{}
    \label{Fig.3a}
\end{subfigure}
\hfill
\begin{subfigure}[t]{0.48\textwidth}
    \centering
    \includegraphics[width=\linewidth]{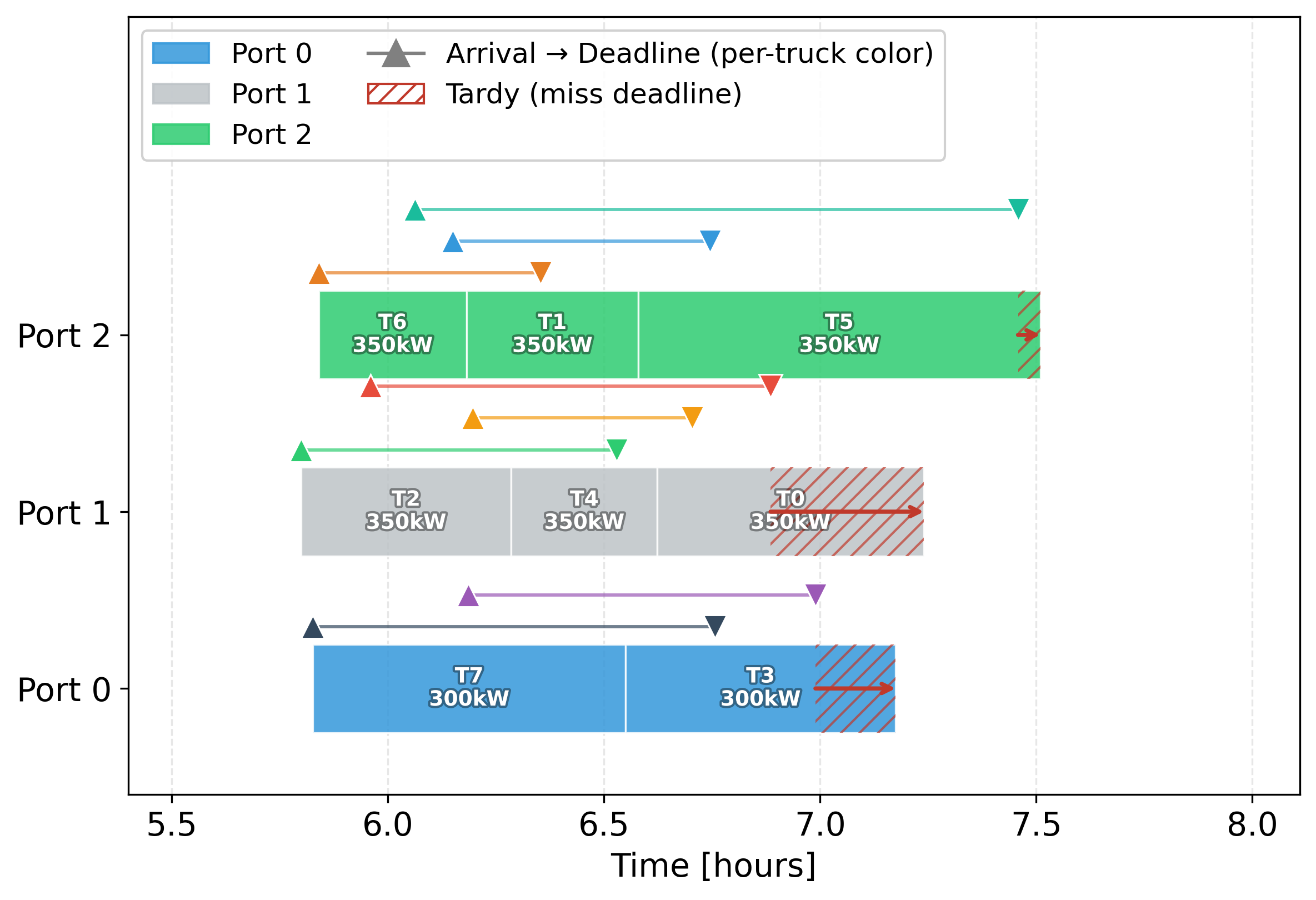}
    \caption{}
    \label{Fig.3b}
\end{subfigure}
\caption{Comparison of the charging schedules under (a) optimal solution, and (b) RO (EDF) solution, for $N\!=\!8$.}
\label{Fig.3}
\end{figure}
We first evaluate the optimality and computational efficiency of the proposed approach by comparing it with the exact solution for small ET fleets, with $N\!=\!4,5,6,7,8$ and $C\!=\!3$. The slack parameter is set to $S\!=\!1.5$, and all trucks are assumed to arrive within a $30$-minute time window. The station-level charging power limit is set as $P^{\max}\!=\!1000$ kW.

Fig.~\ref{Fig.2} shows the total fleet cost under different charging scheduling strategies for $N\!=\!8$. As is shown, the rollout-based approach consistently improves upon the base policies, with RO (EDF) achieving performance identical to the optimal solution. For this scenario, FCFS serves as a worse base policy compared to EDF and SCDF. In Fig.~\ref{Fig.3}, we provide a detailed comparison of the resulting charging schedules under the optimal and RO (EDF) solutions, where $T_i$ denotes the charging duration of truck $i$. The results show that, although the charging order, start times, and effective charging power may differ across trucks, both methods achieve the minimum total cost. These results demonstrate the superior performance of the rollout-based approach and highlight the importance of optimization in reducing the operational cost of ET fleets.

Table~\ref{Table II} further compares the computational time of the optimal and rollout-based solutions, along with the optimality gap between the best rollout solution and the optimal solution. As shown in the table, although the exact DP method guarantees optimality, its computational time grows exponentially with $N$. For a fleet of $8$ ETs at a station with $3$ charging ports, the computation requires approximately $7.82$ hours, making it impractical for real-time planning and unsuitable for large-scale systems. In contrast, the rollout-based approach achieves near-optimal performance across all scenarios while maintaining significantly higher computational efficiency.

\begin{table}[t]
\centering
\fontsize{9}{10}\selectfont
\renewcommand{\arraystretch}{1.1}
\setlength{\tabcolsep}{4pt}
\caption{Comparison of the optimality gap and computational time.}
\label{Table II}
\begin{tabular}{c c c c c c}
\toprule
$N$ 
& Optimal 
& \multicolumn{3}{c}{Rollout [s]} 
& Best Gap \\
\cmidrule(lr){3-5}
 & [s] & RO(FCFS) & RO(EDF) & RO(SCDF) & [\%] \\
\midrule
4 & \cellcolor{purple!7}0.14    & 0.003 & 0.003 & 0.003 & 0.00 \\
5 & \cellcolor{purple!7}2.42    & 0.006 & 0.006 & 0.006 & 0.00 \\
6 & \cellcolor{purple!7}46.56   & 0.010 & 0.010 & 0.010 & 1.94 \\
7 & \cellcolor{purple!7}1376.26 & 0.025 & 0.026 & 0.026 & 8.26 \\
8 & \cellcolor{purple!7}28165.02 & 0.021 & 0.022 & 0.021 & 0.00 \\
\bottomrule
\end{tabular}
\end{table}

\subsubsection{Application to Large Fleet}
Next, we evaluate the performance of the proposed approach for large-scale ET fleets, considering scenarios with $N\!=\!25,50,75,100,125$. The charging station is equipped with $C\!=\!10$ charging ports, and the total power limit is set as $P^{\max}\!=\!3350$ kW. Truck arrival times are randomly generated within the time window of 06:00–12:00. The time-of-use electricity prices specified in Table~\ref{Table I} are adopted, and the slack parameter is set to $S\!=\!2$. The rollout-based approach is compared with the three heuristic methods, and the results are shown in Table~\ref{Table III}.

As we can see from the table, the best-performing rollout method could be based on different base policies. Moreover, as the fleet size increases, the cost reduction achieved by the rollout method relative to the corresponding base policy grows significantly, indicating that the advantage of the rollout-based approach becomes more pronounced in large-scale systems. The results also show that the rollout-based method offers high computational efficiency, making it well suited for efficient fleet operation in large ET fleets. 

\begin{figure}[t]
    \centering
     \includegraphics[width=0.98\linewidth]{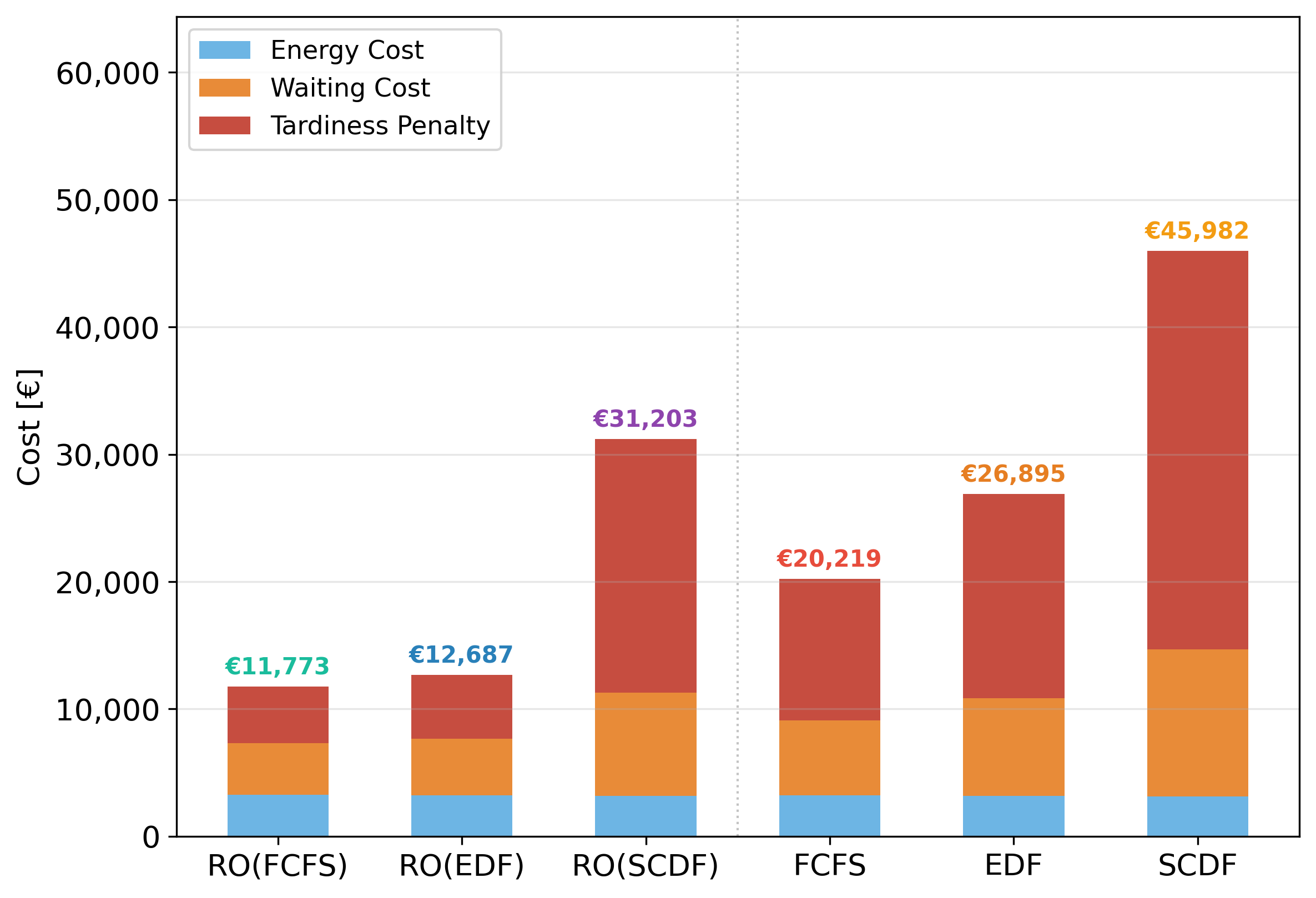}
      \caption{Comparison of the total cost ($N\!=\!100$).}
      \label{Fig.4}
\end{figure}

\begin{figure}[t]
    \centering
     \includegraphics[width=0.97\linewidth]{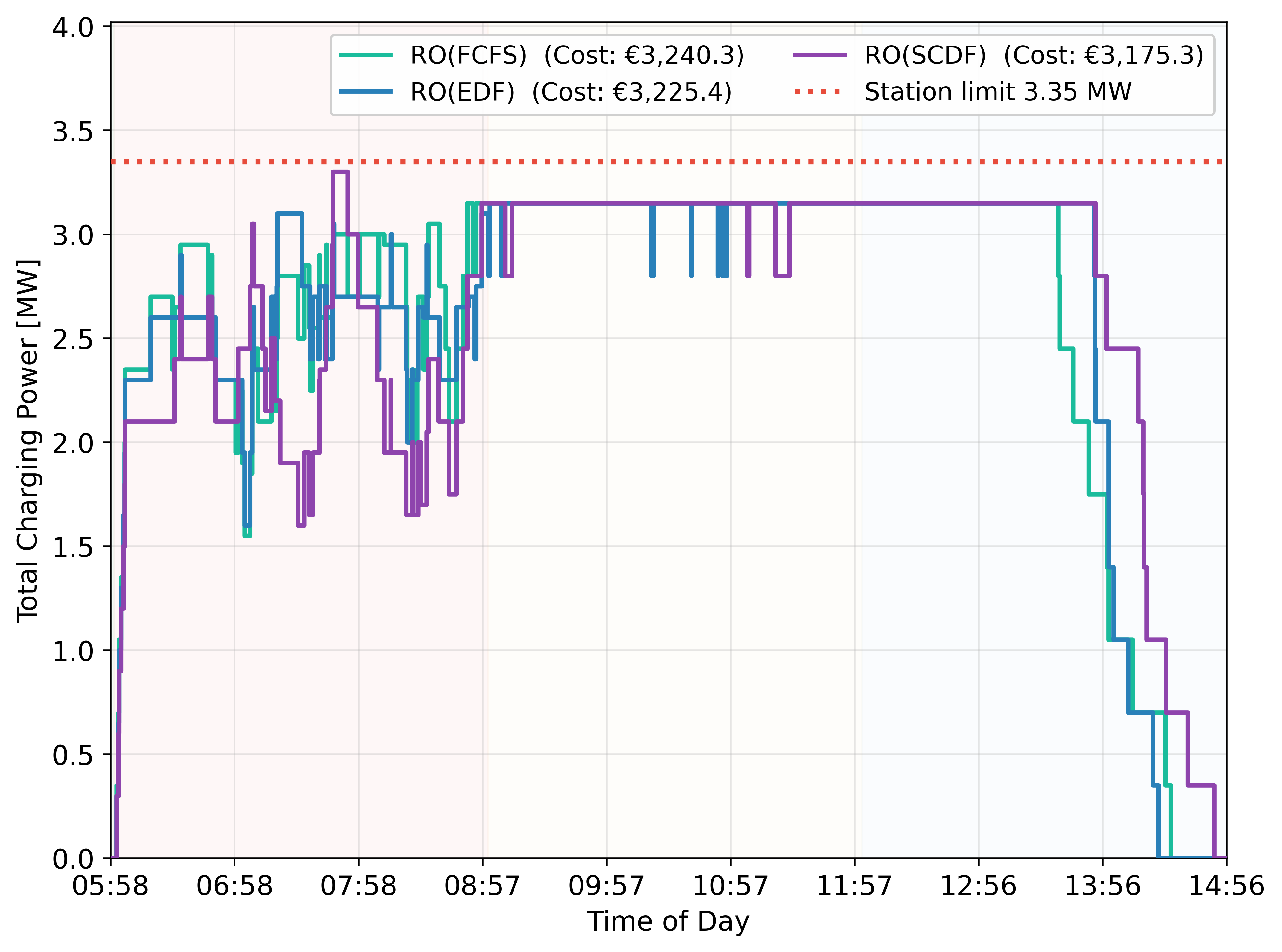}
      \caption{Comparison of the total charging power ($N\!=\!100$).}
      \label{Fig.5}
\end{figure}

Fig.~\ref{Fig.4} presents the total operational cost of the fleet for $N\!=\!100$, comparing the rollout-based policies with the heuristic methods. The results show that RO (FCFS) achieves the lowest total cost, where the cost caused by tardy penalties is reduced significantly compared with other approaches. Moreover, Fig.~\ref{Fig.5} illustrates the corresponding total charging power under different methods. The results show that the station power limit constraint is satisfied in all cases, and that RO (SCDF) attains the lowest charging cost among the rollout-based methods. These results demonstrate the superior performance and high computational efficiency of the proposed approach.

\begin{table*}[t]
\centering
\caption{Best rollout performance compared to the base policy.}
\label{Table III}
\setlength{\tabcolsep}{12pt}
\renewcommand{\arraystretch}{1.1}
\small
\begin{tabular}{l c c c c c}
\toprule
& $N\!=\!25$
& $N\!=\!50$
& $N\!=\!75$
& $N\!=\!100$
& $N\!=\!125$ \\
\midrule
Best Rollout Method
  & RO (FCFS) & RO (FCFS) & RO (FCFS) & RO (FCFS) & RO (EDF) \\
Best Rollout Cost [€]
  & 767.88 & 1{,}789.68 & 2{,}957.73 & 11{,}772.65 & 66{,}881.77 \\
\rowcolor{blue!5}
Cost Reduction vs.\ Base Policy [\%]
  & 0.00 & 2.22 & 31.80 & 41.77 & 35.43 \\
Computation Time [s]
  & 2.38 & 26.98 & 126.52 & 429.65 & 1007.72 \\
\bottomrule
\end{tabular}
\end{table*}

\section{Conclusion}\label{Section VI}
In this paper, we studied the charging scheduling optimization problem for large-scale ET fleets under dedicated charging stations with limited resources, including constrained charging ports and power capacity. The fleet manager optimizes the charging order and schedules for all trucks within the fleet to minimize the total operational cost while meeting the charging demands and delivery deadlines of individual trucks. The problem was formulated as a mixed-integer nonlinear program that captures the coupling between the discrete ordering decisions and continuous charging control. To address the resulting computational challenges, we developed a two-layer optimization framework that decouples the combinatorial ordering decisions from the continuous scheduling problem. A rollout-based DP approach was then proposed to address the problem efficiently. Extensive simulation studies demonstrate the effectiveness and superior performance of the proposed approach. As a future work, we will consider extending the proposed framework to stochastic environments with uncertain arrivals and energy demands. In addition, incorporating learning-based or adaptive base policies could be another direction to further enhance the performance of the rollout-based approach.

\bibliographystyle{IEEEtran}
\bibliography{CDC_2026,IDS_Publications_03272026}
\end{document}